\documentclass[11pt,twoside]{article}
\usepackage{acta-info}
\usepackage{euler}

\usepackage{graphicx}
\usepackage{amsmath}
\usepackage{amsfonts}
\usepackage{amssymb}
\usepackage{latexsym}
\usepackage{epstopdf}
 
\newcommand{\vol}{\textbf{3,} 1 (2011) 5--34}

\newcommand{\amss}{\amssubj{03-2, 68-02, 68Q45, 68R15, 03D15}}
\newcommand{\CRs}{\CRsubj{F.4.3}}
\newcommand{\keyw}{\keywords{primitivity of words, periodicity of words, roots of words and languages, powers of languages, combinatorics on words, Chomsky hierarchy, contextual languages, computational complexity, decidability}}

\newcommand{\N}{{\rm I\!N}}
\newcommand{\zi}{\Sigma }

\date{}

\begin{document}

\title{Primitive words and roots of words}

\maketitle

\oneauthor{%
\href{http://theinf2.informatik.uni-jena.de/People/Gerhard+Lischke.html}{Gerhard LISCHKE} 
}{
\href{http://www.fmi.uni-jena.de/Fakultat.html}{Fakult\"at f\"ur Mathematik und Informatik} \\ \href{http://www.uni-jena.de/}{Friedrich-Schiller-Universit\"at Jena} \\ Ernst-Abbe-Platz 1-4, D-07743 Jena, Germany 
}{
\href{mailto:gerhard.lischke@uni-jena.de}{gerhard.lischke@uni-jena.de}
}

\short{G. Lischke}{Primitive words and roots of words}

\begin{abstract}
In the algebraic theory of codes and formal languages, the set $Q$ of all primitive words over some alphabet $\zi $  has received special interest. With this survey article we give an overview about relevant research to this topic during the last twenty years including own investigations and some new results. In Section 1 after recalling the most important notions from formal language theory we illustrate the connection between coding theory and primitive words by some facts. We define primitive words as words having only a trivial representation as the power of another word. Nonprimitive words (without the empty word) are exactly the periodic words. Every nonempty word is a power of an uniquely determined primitive word which is called the root of the former one. The set of all roots of nonempty words of a language is called the root of the language. The primitive words have interesting combinatorial properties which we consider in Section 2. In Section 3 we investigate the relationship between the set $Q$ of all primitive words over some fixed alphabet and the language classes of the Chomsky Hierarchy and the contextual languages over the same alphabet. The computational complexity of the set $Q$ and of the roots of languages are considered in Section 4. The set of all powers of the same degree of all words from a language is the power of this language. We examine the powers of languages for different sets of exponents, and especially their regularity and context-freeness, in Section 5, and the decidability of appropriate questions in Section 6. Section 7 is dedicated to several generalizations of the notions of periodicity and primitivity of words.
\end{abstract}

\section{Preliminaries}

\subsection{Words and languages}

First, we repeat the most important notions which we will use in our paper.

$\zi $ should be a fixed alphabet, which means, it is a finite and nonempty set of symbols. Mostly, we assume that it is a {\bf nontrivial} alphabet, which means that it has at least two symbols which we will denote by $a$ and $b$, $a\ne b$. $\N = \{0,1,2,3,\ldots \}$ denotes the set of all natural numbers. $\zi ^*$ is the free monoid generated by $\zi $ or the set of all words over $\zi $. The number of letters of a word $p$, with their multiplicities, is the {\bf length} of the word $p$, denoted by $|p|$. If $|p|=n$ and $n=0$, then $p$ is the {\bf empty word}, denoted by $\epsilon $ (in other papers also by $e$ or $\lambda $). The set of words of length $n$ over $\zi $ is denoted by $\zi ^n$. Then $\zi ^* = \bigcup\limits_{n\in \N }\zi ^n$ and $\zi ^0 = \{\epsilon \}$. For the set of nonempty words over $\zi $ we will use the notation $\zi ^+ = \zi ^*\setminus \{\epsilon \}$.

The {\bf concatenation} of two words \,$p = x_1x_2\cdots x_m$ \,and \,$q = y_1y_2\cdots y_n$, \, $x_i,y_j\in \zi $, \,is the word \,$pq = x_1x_2\cdots x_my_1y_2\cdots y_n$. \,We have $|pq| = |p|+|q|$. The {\bf powers} of a word $p\in \zi ^*$ are defined inductively: $p^0=\epsilon $, and $p^n=p^{n-1}p$ for $n\ge 1$. $p^*$ denotes the set $\{p^n: n\in \N \}$, and $p^+=p^*\setminus \{\epsilon \}$.

For $p\in \zi ^*$ and $1\le i\le |p|$, $p[i]$ is the letter at the $i$-th position of $p$. \\ 
Then  \,$p=p[1]p[2]\cdots p[|p|]$.

For words $p,q\in \zi ^*$, $p$ is a {\bf prefix of $q$}, in symbols $p\sqsubseteq q$, if there exists $r\in \zi ^*$ such that $q=pr$. $p$ is a {\bf strict prefix of $q$}, in symbols $p\sqsubset q$, if $p\sqsubseteq q$ and $p\neq q$. $Pr(q) =_{Df} \{p: p\sqsubset q\}$ is the {\bf set of all strict prefixes of $q$} (including $\epsilon $ if $q\neq \epsilon $). \\
$p$ is a {\bf suffix of $q$}, if there exists $r\in \zi ^*$ such that $q=rp$.  

For an arbitrary set $M$, $|M|$ denotes the cardinality of $M$, and ${\cal P}(M)$ denotes the set of all subsets of $M$.

A {\bf language over $\zi $} or a {\bf formal language over $\zi $} is a subset $L$ of $\zi ^*$. \\ $\{L: L\subseteq \zi ^*\} = {\cal P}(\zi ^*)$ is the set of all languages over $\zi $. If $L$ is a nonempty strict subset of $\zi ^*$, $L\subset \zi ^*$, then we call it a nontrivial language.

For languages $L_1$, $L_2$, and $L$ we define: \\ 
$L_1\cdot L_2 = L_1L_2 =_{Df} \{pq: p\in L_1 \wedge  q\in L_2\}$, \\
$L^0 =_{Df} \{\epsilon \}$, \,and \,$L^n =_{Df} L^{n-1}\cdot L$ for $n\ge 1$. \\
If one of $L_1, L_2$ is a one-element set $\{p\}$, then, usually, in $L_1L_2$ we write $p$ instead of $\{p\}$.

Languages can be classified in several ways, for instance according to the Chomsky hierarchy, which we will assume the reader to be familiar with (otherwise, see, for instance, in \cite{8,9,23}). These are the classes of regular, context-free, context-sensitive, and enumerable languages, respectively. Later on we will also consider linear languages and contextual languages and define them in Section 3.

\subsection{Periodic words, primitive words, and codes}

 Two of the fundamental problems of the investigations of words and languages are the questions how a word can be decomposed and whether words are powers of a common word. These occur for instance in coding theory and in the longest repeating segment problem which is one of the most important problems of sequence comparing in molecular biology. The study of primitivity of sequences is often the first step towards the understanding of sequences. 

We will give two definitions of periodic words and primitive words, respectively, and show some connections to coding theory.

\begin{definition}\label{d1} A word $u\in \zi ^+$ is said to be {\bf periodic} if there exists a word $v\in \zi ^*$ and a natural number $n\ge 2$ such that $u=v^n$. If $u\in \zi ^+$ is not periodic, then it is called a {\bf primitive word over $\zi $}.
\end{definition}

Obviously, this definition is equivalent to the following.

\medskip \noindent \textbf{Definition 1$'$} {\em A word $u\in \zi ^+$ is said to be {\bf primitive} if it is not a power of another word, that is, $u=v^n$ with $v\in \zi ^*$ implies $n=1$ and $v=u$. If $u\in \zi ^+$ is not primitive, then it is called a {\bf periodic word over $\zi $}.}

\begin{definition} \label{d2} The set of all periodic words over $\zi $ is denoted by $Per(\zi )$, the set of all primitive words over $\zi $ is denoted by $Q(\zi )$.
\end{definition}

Obviously, \,$Q(\zi ) = \zi ^+\setminus Per(\zi )$.

In the sequel, if $\zi $ is understood, and for simplicity, instead of $Per(\zi )$ and $Q(\zi )$ we will write $Per$ and $Q$, respectively.           

Now we cite some fundamental definitions from coding theory.

\begin{definition} \label{d3} A nonempty set ${\cal C}\subseteq \zi ^*$ is called a {\bf code} if every equation \,$u_1u_2\cdots u_m = v_1v_2\cdots v_n$ \,with $u_i,v_j\in {\cal C}$ for all $i$ and $j$ implies $n=m$ and $u_i=v_i$ for all i. \\
A nonempty set ${\cal C}\subseteq \zi ^*$ is called an {\bf $n$-code} for $n\in \N $, if every nonempty subset of ${\cal C}$ with at most $n$ elements is a code. \, A nonempty set ${\cal C}\subseteq \zi ^+$ is called an {\bf intercode} if there is some $m\ge 1$ such that \, ${\cal C}^{m+1}\cap \zi ^+{\cal C}^m\zi ^+ = \emptyset $.
\end{definition}

Connections to primitive words are stated by the following theorems. 
 
\begin{theorem} \label{t4}
If ${\cal C}\subseteq \zi ^+$ and for all $p,q\in {\cal C}$ with $p\neq q$ holds that $pq\in Q$, then ${\cal C}$ is a 2-code.
\end{theorem}

The proof will be given in Section 2.

\begin{theorem} \label{t5}  If ${\cal C}$ is an intercode, then ${\cal C} \subseteq  Q$.
\end{theorem}

\begin{proof} Assume that ${\cal C}\not\subseteq Q$ is an intercode and ${\cal C}^{m+1}\cap \zi ^+{\cal C}^m\zi ^+ = \emptyset $ for some $m\ge 1$. Then we have a periodic word $u$ in ${\cal C}$ which means $u=v^n \in {\cal C}$ for some $v\in \zi ^+$ and $n\ge 2$. Then \,$u^{m+1} = v^{n(m+1)} = v(v^{nm})v^{n-1} \in {\cal C}^{m+1}\cap \zi ^+{\cal C}^m\zi ^+$, which is a contradiction. 
\end{proof}

\subsection{Roots of words and languages}

 Every nonempty word $p\in \zi ^+$ is either the power of a shorter word $q$ (if it is periodic) or it is not a power of another word (if it is primitive). The shortest word $q$ with this property (in the first case) resp. $p$ itself (in the second case) is called the root of $p$.

\begin{definition}  \label{d6} The {\bf root of a word $p\in \zi ^+$} is the unique primitive word $q$ such that $p=q^n$ for some also unique natural number $n$. It is denoted by $\sqrt{p}$ or $root(p)$. The number $n$ in this equation is called the {\bf degree of $p$}, denoted by $deg(p)$. For a language $L$, \,$\sqrt{L} =_{Df} \{\sqrt{p}: p\in L \wedge p\neq \epsilon \}$ is the {\bf root of $L$}, \,$deg(L) =_{Df} \{deg(p): p\in L \wedge p\neq \epsilon \}$ is the {\bf degree of $L$}.
\end{definition}

\noindent \textbf{Remark.} The uniqueness of root and degree is obvious, a formal proof will be given in Section 2.

\begin{corollary} \label{c7} $p = \sqrt{p}^{deg(p)}$ for each word $p\neq \epsilon $; \, \, $\sqrt{L}\subseteq Q$ for each language $L$; \, \, $\sqrt{\zi ^*} = Q$; \, \, $\sqrt{L}=L$ if and only if $L\subseteq Q$.
\end{corollary}

\vspace*{-3mm}
\section{Primitivity and combinatorics on words} 

 Combinatorics on words is a fundamental part of the theory of words and languages. It is profoundly connected to numerous different fields of mathematics and its applications and it emphasizes the algorithmic nature of problems on words. Its objects are elements from a finitely generated free monoid and therefore combinatorics on words is a part of noncommutative discrete mathematics. For its comprehensive results and its influence to coding theory and primitive words we refer to the textbooks of Yu \cite{28}, Shyr \cite{24}, Lothaire \cite{19}, and to Chapter 6 in \cite{23}. Here we summarize some results from this theory which are important for studying primitive words or which will be used later.

The following theorem was first proved for elements of a free monoid.
 
\begin{theorem}\label{t8} \emph{(Lyndon and Sch\"utzenberger \cite{20})}. If $pq=qp$ for nonempty words $p$ and $q$, then $p$ and $q$ are powers of a common word and therefore $pq$ is not primitive.
\end{theorem}

\begin{figure}
\vspace*{-36pt}
\begin{center}
  \includegraphics[scale=1.3]{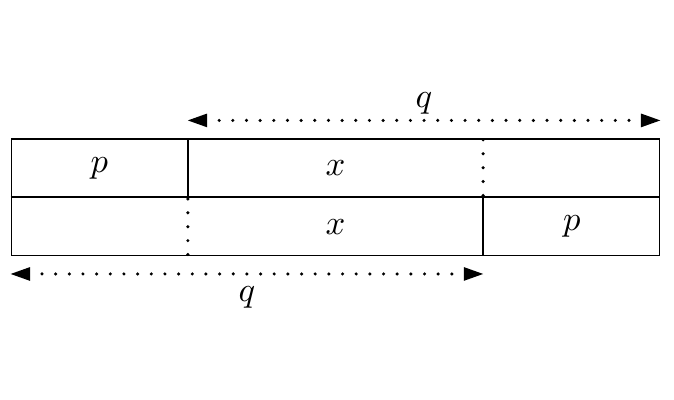}
\end{center}
\vspace*{-48pt}
\caption{To the proof of Theorem \ref{t8}} \label{f1}
\end{figure}

\begin{proof} We prove the theorem by induction on the length of $pq$, which is at least 2. For $|pq|=2$ and $pq=qp$, $p,q\neq \epsilon $, we must have $p=q=a$ for some $a\in \zi $, and the conclusion is true. Now suppose the theorem is true for all $pq$ with $|pq|\le n$ for a fixed $n\ge 2$. Let $|pq|=n+1$, $pq=qp$, $p,q\neq \epsilon $, and, without loss of generality, $|p|\le |q|$. We have a situation as in Figure 1. There must exist $x\in \zi ^*$ such that $q=px=xp$.

Case 1) $x=\epsilon $. Then $p=q$, and the conclusion is true.

Case 2) $x\neq \epsilon $. Since $|px|\le n$, by induction hypothesis $p$ and $x$ are powers of a common word. Then also $q$ is a power of this common word. 

The theorem follows from induction.
\end{proof}

 \begin{corollary} \label{c9}
 $w\notin Q$ if and only if there exist $p,q\in \zi ^+$ such that \\
$w=pq=qp$.
\end{corollary} 

\begin{theorem}  \emph{(Shyr and Thierrin \cite{25})} \label{t10}  For words $p,q\in \zi ^*$, the two-element set $\{p,q\}$ is a code if and only if $pq\neq qp$.
\end{theorem}

\begin{proof} First note, that both statements in the theorem imply, that $p,q\neq \epsilon $ and $p\neq q$. It is trivial that for a code $\{p,q\}$, $pq\neq qp$ must hold. Now we show, that no set $\{p,q\}$ with $pq\neq qp$ can exist which is not a code. Assume the opposite. Then \\
${\cal M} =_{Df} \{\{p,q\}: \,p,q\in \zi ^* \,\wedge  \,pq\neq qp \,\wedge  \,\{p,q\}\mbox{ is not a code}\} \neq  \emptyset $. \\
Let $\{p,q\}\in {\cal M}$ where $|pq|$ is minimal, and let $w$ be a word with minimal length having two different representations over $\{p,q\}$. Then $|w|>2$ and one of the following must be true: \\
either (a) $w=pup=qu'q$ \, or (b) $w=pvq=qv'p$ for some $u,u',v,v'\in \{p,q\}^*$. Because of $p\neq q$, $p\sqsubset q$ or $q\sqsubset p$ must follow. Let us assume that $p\sqsubset q$. For the case $q\sqsubset p$ the proof can be carried out symmetrically. Then from both (a) and (b) it follows that $q=pr=sp$ for some $r,s\in \zi ^+$. We have $|r|=|s|\neq |p|$ (because otherwise $r=s=p$ and $q=pp$), $|pr|<|pq|$, and $pr\neq rp$ (because otherwise $r=s$ and $pq=psp=prp=qp$). With $q=pr$ follows either (a') $pup=pru'pr$ from (a), or (b') $pvpr=prv'p$ from (b). Because of $|pr|<|pq|$, the choice of $\{p,q\}$ having minimal length, and the definition of ${\cal M}$, it must follow that $\{p,r\}$ is a code. But then from both (a') and (b') follows $p=r$, which is a contradiction. Hence ${\cal M}$ must be empty. 
\end{proof}

From the last two theorems we get the following corollary which for its part proves Theorem \ref{t4}.

\begin{corollary} \label{c11} If $pq\in Q$ for words $p,q\neq \epsilon $, then $\{p,q\}$ is a code.
\end{corollary}

Note, that the reversal of this corollary is not true. For example, $\{aba,b\}$ is a code, but $abab\notin Q$.

A weaker variant of the next theorem has been proved also by Lyndon and Sch\"utzenberger \cite{20} for elements of a free monoid. Our proof follows that presented by Lothaire \cite{19}.

\begin{theorem} \emph{(Fine and Wilf \cite{7})} \label{t12} Let $p$ and $q$ be nonempty words, $|p|=n$, $|q|=m$, and $d=gcd(n,m)$ be the greatest common divisor of $n$ and $m$. If $p^i$ and $q^j$ for some $i,j\in \N $ have a common prefix $u$ of length $n+m-d$, then $p$ and $q$ are powers of a common word of length $d$ and therefore $\sqrt{p}=\sqrt{q}$.
\end{theorem}

\begin{proof} Assume that the premises of the theorem are fulfilled and, without loss of generality, $1\le n\le m-1$ (otherwise $n=m=d$ and $p=q=u$). We first assume $d=1$ and show, that $p$ and $q$ are powers of a common letter. \\
Because of $u\sqsubseteq p^i$ and $|u|=m-1+n$ we have \\
(1) \, \, $u[x]=u[x+n]$ \,for $1\le x\le m-1$. \\
Because of $u\sqsubseteq q^j$ we have \\
(2) \, \, $u[y]=u[y+m]$ \,for $1\le y\le n-1$. \\
Because of (1) and $1\le m-n\le m-1$ we have \\
(3) \, \, $u[m]=u[m-n]$. \\
Let now $1\le x\le y\le m-1$ with \, $y-x \equiv  n \mod m$. \, Then we have two cases. 

Case a). $y=x+n\le m-1$, and therefore $u[x]=u[y]$ by (1). 

Case b). $y=x+n-m$. Since $x\le m-1$ we have $x+n-m\le n-1$ and $u[y]=u[x+n-m]=u[x+n]=u[x]$ by (2) and (1).

Hence $u[x]=u[y]$ whenever $1\le x\le y\le m-1$ and $y-x \equiv  n \mod m$. It follows by (1) that $u[x]=u[y]$ whenever $1\le x\le y\le m-1$ and \\
$y-x \equiv k\cdot n \mod m$ \,for some $k\in \N $. Because of $gcd(n,m)=1$, the latter is true if $y-x$ is any value of $\{1,2,\ldots ,m-1\}$. This means, under inclusion of (3), $u[1]=u[2]=\cdots =u[m]$, and $p$ and $q$ are powers of the letter $u[1]$. 

If $d>1$, we argue in exactly the same way assuming $\zi ^d$ instead of $\zi $ as the alphabet. 
\end{proof}

If we assume, $p^i=q^j$ for primitive words $p$ and $q$ and $i,j\in \N \setminus \{0\}$, then by Theorem \ref{t12}, $p$ and $q$ are powers of a common word which can only be $p=q$ itself because of its primitivity. This means the uniqueness of the root of a word which also implies the uniqueness of its degree.

Using Theorem \ref{t12} we can easily prove the next theorem.

\begin{theorem} \emph{(Borwein)} \label{t13} If $w\notin Q$ and $wa\notin Q$, where $w\in \zi ^+$ and $a\in \zi $, then $w\in a^+$.
\end{theorem}

The next theorem belongs to the most frequently referred properties concerning primitive words.
 
\begin{theorem} \emph{(Shyr and Thierrin \cite{26})} \label{t14}  If $u_1u_2\neq \epsilon $ and $u_1u_2=p^i$ for some $p\in Q$, then $u_2u_1=q^i$ for some $q\in Q$. This means, if $u=u_1u_2\neq \epsilon $ and $u'=u_2u_1$, then $deg(u)=deg(u')$, $|\sqrt{u}|=|\sqrt{u'}|$, and therefore $u$ primitive if and only if $u'$ primitive.
\end{theorem}

\begin{proof} Let $u_1u_2=p^i\neq \epsilon $ and $p\in Q$. We consider two cases. 

Case 1). $i=1$, which means, $u_1u_2$ is primitive. Assume that $u_2u_1$ is not primitive and therefore $u_2u_1=q^j$ for some $q\in Q$ and $j\ge 2$. Then $q=q_1q_2\neq \epsilon $ such that $u_2=(q_1q_2)^nq_1$, $u_1=q_2(q_1q_2)^m$, and $j=n+m+1$. It follows that $u_1u_2=(q_2q_1)^{m+n+1}=(q_2q_1)^j$ is not primitive. By this contradiction, $u_2u_1=q^1$ is primitive.

Case 2). $i\ge 2$. Then $p=p_1p_2\neq \epsilon $ such that $u_1=(p_1p_2)^np_1$, $u_2=p_2(p_1p_2)^m$, and $i=n+m+1$. Since $p=p_1p_2$ is primitive, by Case 1 also $q=_{Df}p_2p_1$ is primitive, and $u_2u_1=(p_2p_1)^{m+n+1}=q^i$. 
\end{proof}
The proof of the following theorem, which was first done by Lyndon and Sch\"utzenberger \cite{20} for a free group, is rather difficult and therefore omitted here.

\begin{theorem} \label{t15} If $u^mv^n=w^k \neq \epsilon $ for words $u,v,w\in \zi ^*$ and natural numbers $m,n,k\ge 2$, then $u$,$v$ and $w$ are powers of a common word. \\
We say, that the equation $u^mv^n=w^k$, where $m,n,k\ge 2$ has only trivial solutions.
\end{theorem}

The next two theorems are consequences of Theorem \ref{t15}.
\begin{theorem} \label{t16}  If $p,q\in Q$ with $p\neq q$, then $p^iq^j\in Q$ for all $i,j\ge 2$.
\end{theorem}
This theorem is not true if $i=1$ or $j=1$. For instance, let $p=aba$, $q=baab$, $i=2$, $j=1$.
\begin{theorem} \label{t17}  If $p,q\in Q$ with $p\neq q$ and $i\ge 1$, then there are at most two periodic words in each of the languages $p^iq^*$ and $p^*q^i$.
\end{theorem}
\begin{proof} Assume that there are periodic words in $p^iq^*$, and $p^iq^j$ should be the smallest of them. Then $p^iq^j=r^k$ for some $r\in Q$, $k\ge 2$, $r\neq q$. Let also $p^iq^l=s^m\in Per$, $s\in Q$, $l>j$, $m\ge 2$. Then $s^m=r^kq^{l-j}$, and $l-j=1$ by Theorem \ref{t15}. Therefore at most two words $p^iq^j$ and $p^iq^{j+1}$ in $p^iq^*$ can be periodic. For $p^*q^i$ the proof is done analogously. 
\end{proof}

With essentially more effort, the following can be shown.
\begin{theorem} \emph{(Shyr and Yu \cite{27,28})}  \label{t18} If $p,q\in Q$ with $p\neq q$, then there is at most one periodic word in the language $p^+q^+$.
\end{theorem}

\section{Primitivity and language classes} 

As soon as the set $Q$ of primitive words (over a fixed alphabet $\zi $) was defined, the question arose which is the exact relationship between $Q$ and several known language classes. Here it is important that $\zi $ is a nontrivial alphabet because in the other case all results become trivial or meaningless: If $\zi =\{a\}$ then $Q(\zi )=\zi =\{a\}$ and $Per(\zi )=\{a^n: n\ge 2\}$.

First we will examine the relationship of $Q$ to the classes of the Chomsky hierarchy, and second that to the Marcus contextual languages.

\subsection{Chomsky hierarchy}

 Let us denote by REG, CF and CS the class of all regular languages, the class of all context-free languages and the class of all context-sensitive languages (all over the nontrivial alphabet $\zi $), respectively. It is known from Chomsky Hierarchy that REG $\subset $ CF $\subset $ CS (see, e.g., the textbooks \cite{8,9,23}). It is easy to show that $Q\in \mbox{CS}\setminus \mbox{REG}$, and hence it remains the question whether $Q$ is context-free. Before stating the theorem let us remember that CF is the class of languages which are acceptable by nondeterministic pushdown automata, and CS is the class of languages which are acceptable by nondeterministic linear bounded automata. The latter are Turing machines where the used space on its tapes (this is the number of tape cells touched by the head) is bounded by a constant multiple of the length of the input string. If the accepting automaton is a deterministic one the corresponding language is called a deterministic context-free or a deterministic context-sensitive language, respectively. It can be shown that the deterministic context-free languages are a strict subclass of the context-free languages, whereas it is not yet known whether this inclusion is also strict in the case of context-sensitive languages (This is the famous LBA-problem).

\begin{theorem} \label{t19} $Q$ is deterministic context-sensitive but not regular.
\end{theorem}

\begin{proof} 1. It is easy to see that by a deterministic Turing machine for a given word $u$ can be checked whether it fulfills Definition 1 and thus whether it is not primitive or primitive, and this can be done in space which is a constant multiple of $|u|$.

2. is a corollary from the next theorem.

\begin{theorem} \label{t20} A language containing only a bounded number of primitive words and having an infinite root cannot be regular.
\end{theorem}

 If $Q$ would be regular, then also $\overline{Q} = Per\cup \{\epsilon \}$ would be regular because the class of regular languages is closed under complementation. But $\sqrt{\overline{Q}}=Q$ is infinite and therefore by Theorem \ref{t20} it cannot be regular. 
\end{proof}

\noindent {\bf Proof of Theorem \ref{t20}}.  Let $L$ be a language with an infinite root and a bounded number of primitive words. Further let  \\
$m =_{Df} \max (\{|p|: p\in L\cap Q\}\cup \{0\})$. Assume that $L$ is regular. By the pumping lemma for regular languages, there exists a natural number $n\ge 1$, such that any word $u\in L$ with $|u|\ge n$ has the form $u=xyz$ such that $|xy|\le n$, $y\neq \epsilon $, and $xy^kz\in L$ for all $k\in \N $. Let now $u\in L$ with $|\sqrt{u}|>n$ and $|u|>m$. Then $u=xyz$ such that $1\le |y|\le |xy|\le n$, $z\neq \epsilon $, and $xy^kz\in L$ for all $k\in \N $. By Theorem \ref{t14}, for each $k\ge 1$, $zxy^k$ is periodic (since $|xy^kz|\ge |u|>m$). Let $p =_{Df} \sqrt{zx}$, $i =_{Df} deg(zx)$, and $q =_{Df} \sqrt{y}$. It is $p\neq q$ because otherwise, by Theorem \ref{t14}, $|\sqrt{u}|=|\sqrt{zxy}|=|\sqrt{y}|\le |y|\le n$ contradicting the assumption $|\sqrt{u}|>n$. Then we have infinitely many periodic words in $p^iq^*$ contradicting Theorem \ref{t17}. 
$\square$ 

In 1991 it was conjectured by D\"om\"osi, Horv\'ath and Ito \cite{4} that $Q$ is not context-free. Even though up to now all attempts to prove or disprove this conjecture failed, it is mostly assumed to be true. Some approximations to the solution of this problem will be given with the following theorems.

\begin{theorem} \label{t21}  $Q$ is not deterministic context-free.
\end{theorem}

\begin{proof} We use the fact that the class of deterministic context-free languages is closed under complementation and under intersection with regular sets. Assume that $Q$ is deterministic context-free. Then also \,$\overline{Q}\,\cap \,a^*b^*a^*b^* = \{a^ib^ja^ib^j: i,j\in \N \}$ must be deterministic context-free. But using the pumping lemma for context-free languages, it can be shown that the latter is not even context-free. 
\end{proof}

In the same way (using the pumping lemma for $Per\cap a^*b^*a^*b^*$) it also follows that $Per$ is not context-free.

The next theorem has a rather difficult proof. Therefore and because we will not explain what unambiguity means, we omit the proof.

\begin{theorem} \emph{(Petersen \cite{22})} \label{t22} $Q$ is not an unambigous context-free language.
\end{theorem}

Another interesting language class which is strictly between the context-free and the regular languages is the class LIN of all linear languages.

\begin{definition} \label{d23}  A grammar $G = [N,T,P,S]$ is {\bf linear} if its productions are of the form $A \rightarrow  aB$ or $A \rightarrow  Ba$ or $A \rightarrow  a$, where $a\in T$ and $A,B\in N$. A production of the form $S \rightarrow  \epsilon $ can also be accepted if the start symbol $S$ does not occur in the right-hand side of any production. 

A {\bf linear language} is a language which can be generated by a linear grammar. LIN is the class of all linear languages.
\end{definition}

It can be shown that $\mbox{REG}\subset \mbox{LIN}\subset \mbox{CF}$.

\begin{theorem} \emph{(Horv\'ath \cite{10})} \label{t24}  $Q$ is not a linear language.
\end{theorem}

The proof can be done by using a special pumping lemma for linear languages and will be omitted here.

Let ${\cal L}$ be the union of the classes of linear languages, unambigous context-free languages and deterministic context-free languages. Then ${\cal L}\subset \mbox{CF}$ and, by the former theorems, $Q\notin {\cal L}$. But, whether $Q\in \mbox{CF}$ or not, is still unknown.

\subsection{Contextual languages}

 Though we do not know the exact position of $Q$ in the Chomsky Hierarchy, its position in the system of contextual languages is clear. First, we cite the basic definitions from \cite{21}, see also \cite{15}, and then, after three examples we prove our result.

\begin{definition} \label{d25}  A {\bf (Marcus) contextual grammar} is a structure  $G = $ \linebreak $ [\zi ,A,C,\phi ]$ where $\zi $ is an alphabet, $A$ is a finite subset of $\zi ^*$ (called the set of axioms), $C$ is a finite subset of $\zi ^*\times \zi ^*$ (called the set of contexts), and $\phi $ is a function from $\zi ^*$ into ${\cal P}(C)$ (called the choice function). If $\phi (u)=C$ for every $u\in \zi ^*$ then $G$ is called a {\bf (Marcus) contextual grammar without choice.}
\end{definition} 

With such a grammar the following relations on $\zi ^*$ are associated: For $w,w'\in \zi ^*$, \\
(1) $w \Rightarrow _{ex} w'$ \, if and only if there exists $[p_1,p_2]\in \phi (w)$ such that \\
$w'=p_1wp_2$, \\
(2) $w \Rightarrow _{in} w'$ \, if and only if there exists $w_1,w_2,w_3\in \zi ^*$ and $[p_1,p_2]\in \phi (w_2)$ such that $w=w_1w_2w_3$ and $w'=w_1p_1w_2p_2w_3$. 

$\Rightarrow _{ex}^*$ and $\Rightarrow _{in}^*$ denote the reflexive and transitive closure of these two relations.

\begin{definition} \label{d26}  For a contextual grammar $G = [\zi ,A,C,\phi ]$ (with or without choice), \\
${\cal L}_{ex}(G) =_{Df} \{w: \exists u(u\in A \wedge  u\Rightarrow _{ex}^*w)\}$ is the {\bf external contextual language (with or without choice) generated by} $G$, \\
and \, ${\cal L}_{in}(G) =_{Df} \{w: \exists u(u\in A \wedge  u\Rightarrow _{in}^*w)\}$ is the {\bf internal contextual language (with or without choice) generated by} $G$.
\end{definition}

For every contextual grammar $G = [\zi ,A,C,\phi ]$, \, $A\subseteq {\cal L}_{ex}(G)\subseteq {\cal L}_{in}(G)$ holds.

The above definitions are illustrated by the following examples.  

 {\bf Example 1} \, Let $G = [\zi ,A,C,\phi ]$ be a contextual grammar where $\zi =\{a,b\}$, $A=\{\epsilon ,ab\}$, $C=\{[\epsilon ,\epsilon ],[a,b]\}$, $\phi (\epsilon )=\{[\epsilon ,\epsilon ]\}$, $\phi (ab)=\{[a,b]\}$ and $\phi (w)=\emptyset $ if $w\notin A$. Then ${\cal L}_{ex}(G) = \{\epsilon ,ab,aabb\}$ and \\
${\cal L}_{in}(G) = \{a^nb^n: n\in \N \}$ since \,$ab \Rightarrow _{ex} aabb$, \,$ab \Rightarrow _{in}^* a^nb^n$ \,for every $n\ge 1$, and there does not exist any $w'$ such that $aabb \Rightarrow _{ex} w'$.

{\bf Example 2} \, Let $G = [\zi ,A,C,\phi ]$ be a contextual grammar where \\
$\zi =\{a,b\}$, $A=\{a\}$, $C=\{[\epsilon ,\epsilon ],[\epsilon ,a],[\epsilon ,b]\}$, \\
$\phi (\epsilon )=\{[\epsilon ,\epsilon ]\}$, $\phi (ua)=\{[\epsilon ,b]\}$ for $u\in \zi ^*$ and $\phi (ub)=\{[\epsilon ,a]\}$ for $u\in \zi ^*$. Then ${\cal L}_{ex}(G) = \{a,ab,aba,abab,\ldots \} = a(ba)^*\cup a(ba)^*b$ \,and \,${\cal L}_{in}(G) = a\zi ^*\setminus aa\zi ^*$.

{\bf Example 3} \, Let $u=a_1a_2a_3\cdots $ be an $\omega $-word over a nontrivial alphabet $\zi $ where $a_i\in \zi $ for all $i\ge 1$. Let $G = [\zi ,A,C,\phi ]$ be a contextual grammar where $A=\{\epsilon ,a_1\}$, $C=\{[\epsilon ,\epsilon ]\}\cup \{[\epsilon ,a]: a\in \zi \}$, $\phi (\epsilon )=\{[\epsilon ,\epsilon ]\}$, $\phi (a_1a_2\cdots a_i)=\{[\epsilon ,a_{i+1}]\}$ and $\phi (w)=\emptyset $ if $w$ is not a prefix of $u$. Then ${\cal L}_{ex}(G) = \{\epsilon ,a_1,a_1a_2,a_1a_2a_3,\ldots \} = Pr(u)$ is the set of all prefixes of $u$. Hence, there exist contextual grammars generating languages which are not recursively enumerable.

\begin{theorem} \emph{(Ito \cite{5})} \label{t27} $Q$ is an external contextual language with choice but not an external contextual language without choice or an internal contextual language with or without choice.
\end{theorem}

\begin{proof} 1. Let  $G = [\zi ,\zi ,\{[u,v]: uv\in \zi ^* \wedge  |uv|\le 2\},\phi ]$ be a contextual grammar, where $\phi (w)=\{[u,v]: uv\in \zi ^* \wedge  |uv|\le 2 \wedge  uwv\in Q\}$ for every $w\in \zi ^*$. Then obviously ${\cal L}_{ex}(G)\subseteq Q$. We prove $Q\subseteq {\cal L}_{ex}(G)$ by induction. First we have $\zi \subseteq (\zi \cup \zi ^2)\cap Q\subseteq {\cal L}_{ex}(G)$. Now assume that for a fixed $n\ge 2$ all primitive words $p$ with $|p|\le n$ are in ${\cal L}_{ex}(G)$. Let $u$ be a primitive word of smallest length $\ge n+1$. We have two cases. 

Case a). $u=wx_1x_2$ with $x_1,x_2\in \zi $ and at least one of $w$ and $wx_1$ is in $Q$. Then, by induction hypothesis, $w\in {\cal L}_{ex}(G)$ or $wx_1\in {\cal L}_{ex}(G)$. But then $w\Rightarrow _{ex}wx_1x_2$ or $wx_1\Rightarrow _{ex}wx_1x_2$, and thus $u\in {\cal L}_{ex}(G)$. 

Case b). $u=wx_1x_2$ with $x_1,x_2\in \zi $ and none of $w$ and $wx_1$ is in $Q$. Then, by Theorem \ref{t13}, $w=x_1^{\,i}$ for some $i\ge 1$, hence $u=x_1^{\,i+1}x_2$ with $x_1\neq x_2$, and $x_2\Rightarrow _{ex}x_1x_2\Rightarrow _{ex}x_1x_1x_2\Rightarrow _{ex}\cdots \Rightarrow _{ex}x_1^{\,i+1}x_2$, and therefore $u\in {\cal L}_{ex}(G)$. 

2. Assume that there exists a contextual grammar $G = [\zi ,A,C,\phi ]$ without choice such that $Q={\cal L}_{ex}(G)$. There must be at least one pair $[u,v]\in \phi (w)$ with $uv\neq \epsilon $ for all $w\in \zi ^*$. Let $p=\sqrt{vu}$ and $i=deg(vu)\ge 1$. Because of $p\in Q={\cal L}_{ex}(G)$, also $upv$ would be in ${\cal L}_{ex}(G)$. We have $vup=p^{i+1}$. By Theorem \ref{t14}, $deg(upv)=deg(vup)=i+1\ge 2$ and therefore $upv\notin Q$, which is a contradiction. 

3. Assume $Q={\cal L}_{in}(G)$ for some contextual grammar $G = [\zi ,A,C,\phi ]$ (with or without choice). There must be words $u,v,w\in \zi ^*$ with $uv\neq \epsilon $ and $[u,v]\in \phi (w)$. Let $n=|uwv|$ and $a,b\in \zi $ with $a\neq b$. Then $a^nb^nwa^nb^nuwv\in Q$, but $a^nb^nwa^nb^nuwv\Rightarrow _{in}a^nb^nuwva^nb^nuwv=(a^nb^nuwv)^2\notin Q$, contradicting ${\cal L}_{in}(G)=Q$. 
\end{proof}

\begin{theorem} \label{t28}  $Per$ is not a contextual language of any kind.
\end{theorem}

\begin{proof} Assume $Per = {\cal L}_{ex}(G)$ or $Per = {\cal L}_{in}(G)$ for some contextual grammar \,$G = [\zi ,A,C,\phi ]$ (with or without choice). Let $m$ be a fixed number with \,$m > \max \{|p|: p\in A  \vee  \exists u([p,u]\in C \vee  [u,p]\in C)\}$. Because $a^mb^ma^mb^m\in Per$ we must have $q\in Per$ such that $q\Rightarrow _{ex}a^mb^ma^mb^m$ or $q\Rightarrow _{in}a^mb^ma^mb^m$. \, In the first case, $q=a^ib^ma^mb^j$ with $i<m \vee  j<m$ must follow. But then $q\notin Per$. In the second case, $q=a^ib^ja^kb^l$ with $i<m \vee  j<m \vee  k<m \vee  l<m$ must follow. But then $qq\Rightarrow _{in}a^mb^ma^mb^ma^ib^ja^kb^l\notin Per$ whereas $qq\in Per$. Therefore $Per \neq  {\cal L}_{ex}(G)$ and $Per \neq  {\cal L}_{in}(G)$. 
\end{proof}

\section{Primitivity and complexity} 

 To investigate the computational complexity of $Q$ and that of roots of languages on the one hand is interesting for itself, on the other hand - because $Q=\sqrt{\zi ^*}$ - there was some speculation to get hints for solving the problem of context-freeness of $Q$. First, let us repeat some basic notions from complexity theory.

If ${\cal M}$ is a deterministic Turing machine, then $t_{\cal M}$ is the {\bf time complexity} of ${\cal M}$, defined as follows. If $p\in \zi ^*$, where $\zi $ is the input alphabet of ${\cal M}$, and ${\cal M}$ on input $p$ reaches a final state (we also say ${\cal M}$ {\bf halts on $p$}), then $t_{\cal M}(p)$ is the number of computation steps required by ${\cal M}$ to halt. If ${\cal M}$ does not halt on $p$, then $t_{\cal M}(p)$ is undefined. For natural numbers $n$, \ $t_{\cal M}(n) =_{Df} \max \{t_{\cal M}(p): p\in \zi ^* \wedge \, |p|=n\}$ \, if ${\cal M}$ halts on each word of length $n$. If $t$ is a function over the natural numbers, then TIME($t$) denotes the class of all sets which are accepted by multitape deterministic Turing machines whose time complexity is bounded from above by $t$. Restricting to one-tape machines, the time complexity class is denoted by 1-TIME($t$).

For simplicity, let us write TIME($n^2$) instead of the more exact notation TIME($f$), where $f(n)=n^2$.
 
\begin{theorem} \emph{(Horv\'ath and Kudlek \cite{12})} \label{t29} $Q \in  \mbox{1-TIME}(n^2)$.
\end{theorem}

The proof which will be omitted is based on Corollary 9 and the linear speed-up of time complexity. The latter means that $\mbox{1-TIME}(t') \subseteq  \mbox{1-TIME}(t)$ if $t'\in \mbox{O}(t)$ and $t(n)\ge n^2$ for all $n$. 

The time bound $n^2$ is optimal for accepting $Q$ (or $Per$) by one-tape Turing machines, which is shown by the  next theorem.

\begin{theorem} \emph{(\cite{17})} \label{t30}  For each one-tape Turing machine ${\cal M}$ deciding $Q$, $t_{\cal M}\in \Omega (n^2)$ must hold. The latter means: \\
$\exists c\exists n_0(c>0\wedge n_0\in \N \wedge \forall n(n\ge n_0\rightarrow t_{\cal M}(n)\ge c\cdot n^2))$.
\end{theorem}

The proof which will be omitted also, uses the for complexity theorists well-known method of counting the crossing sequences.

Now we turn to the relationship between the complexity of a language and that of its root. It turns out that there is no general relation, even more, there can be an arbitrary large gap between the complexity of a language and that of its root. 

\begin{theorem} \emph{(\cite{17,16})} \label{t31}  Let $t$ and $f$ be arbitrary total functions over $\N $ such that $t\in \omega (n)$ is monotone nondecreasing and $f$ is monotone nondecreasing, unbounded, and time constructible. Then there exists a language $L$ such that $L\in \mbox{1-TIME}(\mbox{O}(t))$ but \,$\sqrt{L}\notin \mbox{TIME}(f)$.
\end{theorem}

Instead of the proof which is a little bit complicated we only explain the notions occuring in the theorem. $t\in \omega (n)$ means $\lim\limits_{n\to\infty}\frac{n}{t(n)}=0$. A time constructible function is a function $f$ for which there is a Turing machine halting in exactly $f(n)$ steps on every input of length $n$ for each $n\in \N$. One can show that the most common functions have these properties. Finally, \\
$\mbox{1-TIME}(\mbox{O}(t)) = \bigcup \{\mbox{1-TIME}(t'): \exists c\exists n_0(c>0 \wedge  n_0\in \N  \wedge  \forall n(n\ge n_0 \rightarrow  t'(n)\le c\cdot t(n)))\}$.

Let us still remark, that from Theorem \ref{t31} we can deduce that there exist regular languages the roots of which are not even context-sensitive, see \cite{15,16}.

\section{Powers of languages}

 In arithmetics powers in some sense are counterparts to roots. Also for formal languages we can define powers, and also here we shall establish some connections to roots. For the first time, the power $pow(L)$ of a language $L$ was defined by Calbrix and Nivat in \cite{3} in connection with the study of properties of period and prefix languages of $\omega $-languages. They also raised the problem to characterize those regular languages whose powers are also regular, and to decide the problem whether a given regular language has this property. Cachat \cite{2} gave a partial solution to this problem showing that for a regular language $L$ over a one-letter alphabet, it is decidable whether $pow(L)$ is regular. Also he suggested to consider as the set of exponents not only the whole set $\N $ of natural numbers but also an arbitrary regular set of natural numbers. This suggestion was taken up in \cite{13} with the next definition.

\begin{definition} \label{d32}  For a language $L\subseteq \zi ^*$ and a natural number $k\in \N $, \\
$L^{(k)} =_{Df} \{p^k: p\in L\}$. \, For $H\subseteq \N $, \\
$pow_H(L) =_{Df} \bigcup\limits_{k\in H}L^{(k)} = \{p^k: p\in L \wedge  k\in H\}$ \,is the {\bf $H$-power} of $L$. \\
Instead of $pow_H(L)$ we also write $L^{(H)}$, and also it is usual to write $pow(L)$ instead of $pow_{\N }(L) = L^{(\N )}$.
\end{definition}

Note the difference between $L^{(k)}$ and $L^k$. For instance, if $L = \{a,b\}$ then \\
$L^{(2)} = \{aa,bb\}$,\, $L^2 = \{aa,ab,ba,bb\}$\, and \,$L^{(\N )} = a^*\cup b^*$.

We say that a set $H$ of natural numbers has some language theoretical property if the corresponding one-symbol language $\{a^k: k\in H\} = \{a\}^{(H)}$ which is isomorphic to $H$ has this property.

It is easy to see that every regular power of a regular language is context-sensitive. More generally, we have the following theorem.

\begin{theorem} \emph{(\cite{13})} \label{t33} If $H\subseteq \N $ is context-sensitive and $L\in \mbox{CS}$ then also $pow_H(L) = L^{(H)}$ is context-sensitive.
\end{theorem}

\begin{proof} Let $L\subseteq \zi ^*$ be context-sensitive and also $H\subseteq \N $ be context-sensitive. By the following algorithm, for a given word $u\in \zi ^*$ we can decide whether $u\in L^{(H)}$. 
\begin{tabbing}
123 \= {\bf if} \= {\bf then} \= {\bf for} \= {\bf do if} \= \kill
1 \> {\bf if} $(u\in L \wedge  1\in H) \vee  (u=\epsilon  \wedge  0\in H)$ \\
2 \> \> {\bf then return} ``$u$ is in $L^{(H)}$'' \\
3 \> \> {\bf else} \ compute $p=\sqrt{u}$ and $d=deg(u)$ \\
4 \> \> \> {\bf for} $i\leftarrow 1$ {\bf to} $\lfloor \frac{d}{2}\rfloor $ \\
5 \> \> \> \> {\bf do if} $p^i\in L \wedge  \frac{d}{i}\in H$ \\
6 \> \> \> \> \> {\bf then return} ``$u$ is in $L^{(H)}$'' \\
7 \> \> \> {\bf return} ``$u$ is not in $L^{(H)}$'' 
\end{tabbing} 
$\lfloor \frac{d}{2}\rfloor$ in line 4 is $\frac{d}{2}$ if $d$ is even, and $\frac{d-1}{2}$ if $d$ is odd. Each step of the algorithm can be done by a linear bounded automaton or by a Turing machine where the used space is bounded by a constant multiple of $|u|$. Crucial for this are that $|p|\le |u|$, $d\le |u|$, and the decisions in line 1 and in line 5 can also be done by a linear bounded automaton with this boundary, because $L$ and $H$ are context-sensitive and therefore acceptable by linear bounded automata. 
\end{proof}

The last theorem raises the question whether and when $L^{(H)}$ is in a smaller class of the Chomsky hierarchy, especially if $L$ is regular. This essentially depends on whether the root of $L$ is finite or not. Therefore we will introduce the notions FR for the class of all regular languages $L$ such that $\sqrt{L}$ is finite, and $\mbox{IR} =_{Df} \mbox{REG}\setminus \mbox{FR}$ for the class of all regular languages $L$ such that $\sqrt{L}$ is infinite.

\begin{theorem} \emph{(\cite{13})} \label{t34}  The class FR of regular sets having a finite root is closed under the power with finite sets.
\end{theorem}

\begin{proof} Let $L$ be a regular language with a finite root $\{p_1,\ldots ,p_k\}$ and $\epsilon \notin L$, and let $L_i =_{Df} L\cap p_i^{\,*}$ for each $i\in \{1,\ldots ,k\}$. Since $L_i\subseteq p_i^{\,*}$ and $L_i\in \mbox{REG}$, $L_i$ is isomorphic to a regular set $M_i$ of natural numbers, namely $M_i=deg(L_i)$. For each $n\in \N $, $M_i\cdot n =_{Df} \{m\cdot n: m\in M_i\}$ is regular too. Therefore, for a finite set $H\subseteq \N $, also $\bigcup\limits_{n\in H}M_i\cdot n$ \,is regular which is isomorphic to $L_i^{(H)}$. Then $L^{(H)} = \bigcup\limits_{i=1}^{k}L_i^{(H)}$ is regular, and $\sqrt{L^{(H)}} = \sqrt{L}$ \,is finite. If the empty word is in the language then, because of $pow_H(L\cup \{\epsilon \}) = pow_H(L)\cup \{\epsilon \}$ we get the same result. 
\end{proof}

If $H$ is infinite then $pow_H(L)$ may be nonregular and even non-context-free. This is true even in the case of a one-letter alphabet where the root of each nonempty set (except $\{\epsilon \}$) has exactly one element. This is illustrated by the following example. \\
Let $L = \{a^{2m+3}: m\in \N \}$. Then $L\in \mbox{FR}$ \, but \\
$L^{(\N )} = \{a^k: k\in \N \setminus \{2^m: m\ge 0\}\} \notin \mbox{CF}$. \\
Therefore it remains a problem to characterize those regular sets $L$ with finite roots where $pow_H(L)$ is regular for any (maybe regular) set $H$. 

Our next theorem shows that the powers of arbitrary (not necessarily regular) languages which have infinite roots are not regular, even more, they are not even context-free, if the exponent set is an arbitrary set of natural numbers containing at least one element which is greater than 2 and does not contain the number 1, or some other properties are fulfilled.

\begin{theorem} \emph{(\cite{13})} \label{t35} For every language $L$ which has an infinite root and for every set $H\subseteq \N $ containing at least one number greater than 2, $pow_H(L)$ is not context-free if one of the following conditions is true: 
\begin{tabbing}
(b) \, $\sqrt{L}\in \mbox{REG}$,1234 \= \kill
(a) \, $1\notin H$, \> (c) \, $L\cap \sqrt{L}\in \mbox{REG}$, \\
(b) \, $\sqrt{L}\in \mbox{REG}$, \> (d) \, $L\in \mbox{REG}$ and $\sqrt{L^{(H)}\setminus L}$ is infinite.
\end{tabbing}
\end{theorem}

\begin{proof} Let $L\subseteq \zi ^*$ be a language such that $\sqrt{L}$ is infinite, and let $H\subseteq \N $ with $H\setminus \{0,1,2\} \neq  \emptyset $. We define \\
$$
L' =_{Df} \left\{
\begin{array}{lcc}
pow_H(L) &\mbox{if} & (a) \mbox{ is true, } \\
pow_H(L)\setminus\sqrt{L} & \mbox{if} & (b) \mbox{ is true, } \\
pow_H(L)\setminus(L\cap\sqrt{L}) & \mbox{if} & (c) \mbox{ is true, } \\
pow_H(L)\setminus L & \mbox{if} & (d) \mbox{ is true. } 
\end{array}
\right.
$$ 
\noindent If more than one of the conditions $(a),\,(b),\,(c),\,(d)$ are true simultaneously, then it doesn't matter which of the appropriate lines in the definition of $L'$ we choose. It is important that in each case, $\sqrt{L'}$ is infinite, there is no primitive word in $L'$ and, if $pow_H(L)$ was context-free then also $L'$ would be context-free. But we show that the latter is not true. \\
Assume that $L'$ is context-free, and let $n\ge 3$ be a fixed number from $H$. By the pumping lemma for context-free languages, there exists a natural number $m$ such that every $z\in L'$ with $|z|>m$ is of the form $w_1w_2w_3w_4w_5$ where: $w_2w_4\neq \epsilon $, $|w_2w_3w_4|<m$, and $w_1w_2^{\,i}w_3w_4^{\,i}w_5\in L'$ for all $i\in \N $. \\
Now let $z\in L'$ with $deg(z)\ge n$ and $|\sqrt{z}|>2m$ which exists because $\sqrt{L'}$ is infinite. Let \,$p =_{Df} \sqrt{z}$ \,and \,$k =_{Df} deg(z)$.\, Then $|z|=k\cdot |p|>2km$. By the pumping lemma, $z = p^k = w_1w_2w_3w_4w_5$ where $w_2w_4\neq \epsilon $, $|w_2w_3w_4|<m<\frac{|p|}{2}$, and $w_1w_2^{\,i}w_3w_4^{\,i}w_5\in L'$ for each $i\in \N $. Especially, for $i=0$, $x =_{Df} w_1w_3w_5 \in  L'$ and therefore $x$ is nonprimitive. Now let $z' =_{Df} w_5w_1w_2w_3w_4$, $q =_{Df} \sqrt{z'}$, $x' =_{Df} w_5w_1w_3$, and $s =_{Df} \sqrt{x'}$. By Theorem \ref{t14} we have $deg(z')=deg(z)=k$ and $x'$ nonprimitive, therefore $|q|=|p|>2m$ and $|s|\le \frac{|x'|}{2}$. It follows $z'=q^k$ and $x'=q^{k-1}q'$ for some word $q'$ with $\frac{|q|}{2}<|q'|<|q|$ (because of $0<|w_2w_4|\le |w_2w_3w_4|<\frac{|q|}{2}$). The words $z'$ and $x'$ which are powers of $q$ and $s$, respectively, have a common prefix $w_5w_1$ of length $|z|-|w_2w_3w_4|>k\cdot |q|-\frac{|q|}{2}$. Because of $|s|\le \frac{|x'|}{2}<\frac{k}{2}\cdot |q|$ and $k\ge 3$, we have $|q|+|s|<(\frac{k}{2}+1)|q|\le (k-\frac{1}{2})|q|$, and therefore $q=s$ by Theorem \ref{t12}. But then $x'=s^{k-1}q'$ with $0<|q'|<|s|$ which contradicts $\sqrt{x'}=s$. 
\end{proof} 

It remains open whether the $H$-power of a regular language is regular or context-free or neither, if $H=\N $ or $H\subseteq \{0,1,2\}$. First, we consider the exceptions 0, 1, and 2 where we find out a different behavior.

\begin{theorem} \emph{(\cite{13})} \label{t36} (i) For each $L\in \mbox{REG}$ and $H\subseteq \{0,1\}$, $L^{(H)}\in \mbox{REG}$. \\
(ii) For each $L\in \mbox{FR}$, $L^{(2)}\in \mbox{FR}$. \\
(iii) For each $L\in \mbox{IR}$, $L^{(2)}\notin \mbox{REG}$.
\end{theorem}

\begin{proof} (i) is trivial, (ii) follows from Theorem \ref{t34}. (iii) follows from Theorem \ref{t20}. 
\end{proof}

A set $pow_{\{2\}}(L) = L^{(2)}$ we call also the {\bf square} of $L$. Because of the former theorem, only the squares of regular languages with infinite roots remain for interest. In contrast to the former results where the power of a regular set either is regular again or not context-free, this is not true for the squares. It is illustrated by the following examples: 

Let $L_1 =_{Df} a\cdot \{b\}^*$ and $L_2 =_{Df} \{a,b\}^*$. Then both $L_1$ and $L_2$ are regular with infinite roots, but $L_1^{(2)}\in \mbox{CF}$ and $L_2^{(2)}\notin \mbox{CF}$. 

To characterize those regular languages whose squares are context-free we introduce the following notion.

\begin{definition} \label{d37} Let $p\in Q$ and $w,w'\in \zi ^*$ such that $p$ is not a suffix of $w$ and $w'w\notin p^+$. The sets \, $wp^*w'$ \, and \, $p^*wp^*$ \, are called {\bf inserted iterations of the primitive word $p$}. The words $p$, $w$, $w'$ are called the {\bf modules} of $wp^*w'$, and $p$, $w$ are called the {\bf modules} of $p^*wp^*$. A {\bf FIP-set} is a finite union $L_1\cup \ldots \cup L_n$ of inserted iterations of primitive words. The sets $L_1,\ldots ,L_n$ are also called the {\bf components} of the FIP-set.
\end{definition}

Using this notion we can give the following reformulation and simplification of a theorem by Ito and Katsura from 1991 (see \cite{14}) which has a rather difficult proof.

\begin{theorem} \label{t38} If $L^{(2)}\in \mbox{CF}$ and $L^{(2)}\subseteq Q^{(2)}$ then $L$ must be a subset of a FIP-set.
\end{theorem}

Using this theorem and the proof idea from Theorem \ref{t35} we can show the following characterization.

\begin{theorem} \emph{(\cite{18})} \label{t39}  For a regular language $L$, $L^{(2)}$ is context-free if and only if $L$ is a subset of a FIP-set.
\end{theorem}

\begin{proof} We show here only one direction. Let $L$ be regular and $L^{(2)}\in \mbox{CF}$. We consider three cases.
Case a). $L\in \mbox{FR}$. Let $\sqrt{L}=\{p_1,\ldots ,p_n\}$. Then $L\subseteq p_1^{\,*}\cup \cdots \cup p_n^{\,*}$ and $p_1^{\,*}\cup \cdots \cup p_n^{\,*}$ is a FIP-set. 

Case b). $L\in \mbox{IR}$ and $\sqrt{L\cap Per}$ is infinite. This means, $L$ has infinitely many periodic words with altogether infinitely many roots of unbounded lengths. Then $L^{(2)}$ contains words $z$ with $|\sqrt{z}|>2m$ for arbitrary $m$ and $ deg(z)\ge 4$. If $L^{(2)}$ would be context-free then we would get the same contradiction as in the proof of Theorem \ref{t35}. Therefore case b) cannot occur.

Case c). $L\in \mbox{IR}$ and $\sqrt{L\cap Per}$ is finite. Let $L_1 =_{Df} L\cap Q$, \,$L_2 =_{Df} L\cap \overline{Q}$, \,and \,$\sqrt{L_2} = \{p_1,\ldots ,p_k\}$. Then \,$L = L_1\cup L_2$, \,$L_1\cap L_2 = \emptyset $, \,and \\
$L_2 = ((p_1^{\,*}\cup \cdots \cup p_k^{\,*})\setminus \{p_1,\ldots ,p_k\})\cap L$ \,is in FR. Therefore also $L_2^{(2)}\in \mbox{FR}$ by Theorem \ref{t36}, and $L_1^{(2)}\in \mbox{CF}$ because $L^{(2)} = L_1^{(2)}\cup L_2^{(2)} \in  \mbox{CF}$. \,We have $L_1^{(2)}\subseteq Q^{(2)}$, and by Theorem \ref{t38} follows that $L_1$ is a subset of a FIP-set. $L_2$ is a subset of a FIP-set by case a), and so is $L=L_1\cup L_2$. 
\end{proof}

Now it is easy to clarify the situation for the $n$-th power of a regular or even context-free set for an arbitrary natural number $n$, where it is trivial that $L^{(0)}=\{\epsilon \}$, \,$L^{(1)}=L$.

\begin{theorem} \emph{(\cite{18})} \label{t40} For an arbitrary context-free language $L$ and a natural number $n\ge 2$, if $L^{(n)}$ is context-free, then either $n\ge 3$ and $L\in \mbox{FR}$ or $n=2$ and $L\cap Per \in \mbox{FR}$.
\end{theorem}

\begin{proof} If $n\ge 3$ and $\sqrt{L}$ is infinite then $L^{(n)}\notin \mbox{CF}$ by Theorem \ref{t35}. It is well-known that every context-free language over a single-letter alphabet is regular. Using this fact it is easy to show that every context-free language with finite root is regular too. Therefore, if $\sqrt{L}$ is finite and $L\in \mbox{CF}$ then $L\in \mbox{FR}$, and $L^{(n)}\in \mbox{FR}$ by Theorem \ref{t34}. If $n=2$, $L^{(n)}\in \mbox{CF}$ and $\sqrt{L}$ is infinite, then $L\cap Per \in \mbox{FR}$ must be true by the proof of Theorem \ref{t39}.
\end{proof}

Now we consider the full power $pow(L) = pow_{\N }(L)$ for a regular language $L$.
 
\begin{theorem} \emph{(Fazekas \cite{6})} \label{t41}  For a regular language $L$, $pow(L)$ is regular if and only if $pow(L)\setminus L \in \mbox{FR}$.
\end{theorem}

\begin{proof} If $pow(L)\setminus L \in \mbox{FR} \subseteq  \mbox{REG}$ then $(pow(L)\setminus L)\cup L = pow(L) \in  \mbox{REG}$ because the class of regular languages is closed under union. For the opposite direction assume $pow(L)\in \mbox{REG}$. Then also $L' =_{Df} pow(L)\setminus L$ is regular because the class of regular languages is closed under difference of two sets. There are no primitive words in $L'$ and therefore, by Theorem \ref{t20}, it must have a finite root. 
\end{proof}

\section{Decidability questions} 

 Questions about the decidability of several properties of sets or decidability of problems belong to the most important questions in (theoretical) computer science. Here we consider the decidability of properties of languages regarding their roots and powers. We will cite the most important theorems in chronological order of their proofs but we omit the proofs because of their complexity.
 
\begin{theorem} \emph{(Horv\'ath and Ito \cite{11})} \label{t42} For a context-free language $L$ it is decidable whether $\sqrt{L}$ is finite.
\end{theorem}

\begin{theorem} \emph{(Cachat \cite{2})} \label{t43} For a regular or context-free language $L$ over single-letter alphabet it is decidable whether $pow(L)$ is regular.
\end{theorem}

Using Cachat's algorithm, Horv\'ath showed (but not yet published) the following.

\begin{theorem} \emph{(Horv\'ath)} \label{t44} For a regular or context-free language $L$ with finite root it is decidable whether $pow(L)$ is regular.
\end{theorem}

\textbf{Remark.} Since the context-free languages with finite root are exactly the languages in FR (Remark in the proof of Theorem \ref{t40}), it doesn't matter whether we speak of regularity or context-freeness in the last theorems.

Remarkable in this connection is also the only negative decidability result by Bordihn.

\begin{theorem} \emph{(Bordihn \cite{1})} \label{t45} For a context-free language $L$ with infinite root it is not decidable whether $pow(L)$ is context-free.
\end{theorem}

The problem of Calbrix and Nivat \cite{3} and the open question of Cachat \cite{2} for languages over any finite alphabet and almost any sets of exponents, but not for all, was answered in \cite{13}. Especially the regularity of $pow(L)$ for a regular set $L$ remained open, but it was conjectured that the latter is decidable. Using these papers, finally Fazekas \cite{6} could prove this conjecture.

\begin{theorem} \emph{(Fazekas \cite{6})} \label{t46} For a regular language $L$ it is decidable whether $pow(L)$ is regular.
\end{theorem}

Finally, we look at the squares of regular and context-free languages.

\begin{theorem} \emph{(\cite{18})} \label{t47} For a regular language $L$ it is decidable whether $L^{(2)}$ is regular or context-free or none of them.
\end{theorem}

\begin{proof} Let $L$ be a regular language generated by a right-linear grammar $G = [\zi ,N,S,R]$ and let $m=|N|+1$. By Theorem \ref{t36}, $L^{(2)}$ is regular if and only if $\sqrt{L}$ is finite. The latter is decidable by Theorem \ref{t42}. If $\sqrt{L}$ is infinite then by Theorem \ref{t39}, $L^{(2)}$ is context-free if and only if $L$ is a subset of a FIP-set. If $L$ is a subset of a FIP-set then we can show that there exists a FIP-set $F$ such that $L\subseteq F$ and all modules of all components of $F$ have lengths smaller than $m$. Thus there are only finitely many words which can be modules and only finitely many inserted iterations of primitive words having these modules. The latter can be effectively computed. Let $L_1,\ldots ,L_n$ be all these inserted iterations of primitive words. Then $L^{(2)}$ is context-free if and only if \,$L \subseteq  L_1\cup \cdots \cup L_n$ \,which is equivalent to \,$L\cap \overline{(L_1\cup \cdots \cup L_n)} = \emptyset $. \,The latter is decidable for regular languages $L$ and $L_1,\ldots ,L_n$. 
\end{proof}

\section{Generalizations of periodicity and primitivity} 

 If $u$ is a periodic word then we have a strict prefix $v$ of $u$ such that $u$ is exhausted by concatenation of two or more copies of $v$, $u=v^n$, $n\ge 2$ (see Figure \ref{f3}). But it could be that such an exhaustion is not completely possible, there may remain a strict prefix of $v$ and the rest of $v$ overhangs $u$, i.e. $u=v^nv'$, $n\ge 2$, $v'\sqsubset v$ (see Figure \ref{f4}). In such case we call $u$ to be semi-periodic. A third possibility is to exhaust $u$ by concatenation of two or more copies of $v$ where several consecutive copies may overlap (see Figure \ref{f5}). In this case we speak about quasi-periodic words. If a nonempty word is not periodic, semi-periodic, or quasi-periodic, respectively, we call it a primitive, strongly primitive, or hyperprimitive word, respectively. Of course, periodic and primitive words are those we considered before in this paper. Finally, we can combine the possibilities to get three further types which we will summarize in the forthcoming Definition \ref{d49}. Before doing so, we give a formal definition of concatenation with overlaps. All these generalizations have been introduced and detailed investigated in \cite{15}. Most of the material in this section is taken from there.

\begin{figure}
\begin{center}
   \includegraphics[scale=0.6]{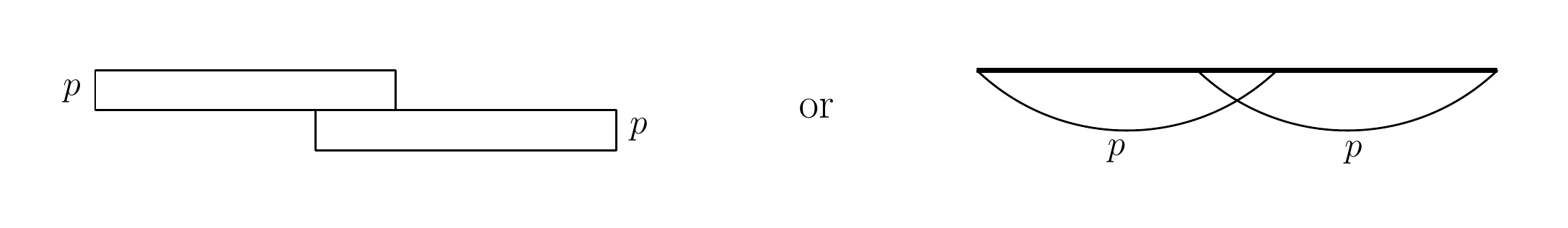}
\end{center}
\vspace*{-24pt}
\caption{Concatenation with overlap} \label{f2}
\end{figure}
 
\begin{definition} \label{d48} For $p,q\in \zi ^*$, we define \\ 
$p\otimes q =_{Df} \{w_1w_2w_3: \,w_1w_3\neq \epsilon   \,\wedge  \,w_1w_2=p \,\wedge  \,w_2w_3=q\}$, \\
$p^{\otimes 0} =_{Df} \{\epsilon \}$, \, $p^{\otimes k+1} =_{Df} \bigcup \{w\otimes p: w\in p^{\otimes k}\}$ \, for $k\in \N $, \\
$A\otimes B =_{Df} \bigcup \{p\otimes q: p\in A \,\wedge \,q\in B\}$ \, for sets $A,B\subseteq \zi ^*$.
\end{definition}

The following example shows that in general, $p\otimes q$ is a set of words: \\
Let $p = aabaa$. Then $p\otimes p = p^{\otimes 2} = \{aabaaaabaa, aabaaabaa, aabaabaa\}$.
We can illustrate this by Figure \ref{f2}.

\medskip
In the following definition we repeat our Definitions \ref{d1} and \ref{d2} and give the generalizations suggested above.

\newpage
\begin{definition} \label{d49} $\,$ \\
\noindent \begin{tabular}{rcl} 
$Per$ & $=_{Df}$ & $\{u: \exists v\exists n(v\sqsubset u \wedge  n\ge 2 \wedge  u=v^n)\}$ \, \, \, is the set of \\
& & {\bf periodic} words. \\
$Q$ & $=_{Df}$ & $\zi ^+\setminus Per$ \, \, \, is the set of \,{\bf primitive} words. \\[2mm]
$SPer$ & $=_{Df}$ & $\{u: \exists v\exists n(v\sqsubset u \wedge  n\ge 2 \wedge  u\in v^n\cdot Pr(v))\}$ \, \, \, is the \\
& & set of \,{\bf semi-periodic} words. \\
$SQ$ & $=_{Df}$ & $\zi ^+\setminus SPer$ \, \, \, is the set of \,{\bf strongly primitive} words.  \\[2mm]
$QPer$ & $=_{Df}$ & $\{u: \exists v\exists n(v\sqsubset u \wedge  n\ge 2 \wedge  u\in v^{\otimes n})\}$ \, \, \, is the set of \\
& & {\bf quasi-periodic} words. \\
$HQ$ & $=_{Df}$ & $\zi ^+\setminus QPer$ \, \, \, is the set of \,{\bf hyperprimitive} words. \\[2mm]  
$PSPer$ & $=_{Df}$ & $\{u: \exists v\exists n(v\sqsubset u \wedge  n\ge 2 \wedge  u\in \{v^n\}\otimes Pr(v))\}$ \, \, \, is the \\
& & set of \,{\bf pre-periodic} words. \\
$SSQ$ & $=_{Df}$ & $\zi ^+\setminus PSPer$ \, \, \, is the set of \,{\bf super strongly primitive} \\
& & words. \\[2mm]
$SQPer$ & $=_{Df}$ & $\{u: \exists v\exists n(v\sqsubset u \wedge  n\ge 2 \wedge  u\in v^{\otimes n}\cdot Pr(v))\}$ \, \,\,is the \\
& & set of \,{\bf semi-quasi-periodic} words. \\
$SHQ$ & $=_{Df}$ & $\zi ^+\setminus SQPer$ \, \, \, is the set of \,{\bf strongly hyperprimitive} \\
& & words.  \\[2mm]
$QQPer$ & $=_{Df}$ & $\{u: \exists v\exists n(v\sqsubset u \wedge  n\ge 2 \wedge  u\in v^{\otimes n}\otimes Pr(v))\}$ \,\, is the \\
& & set of \,{\bf quasi-quasi-periodic} words. \\
$HHQ$ & $=_{Df}$ & $\zi ^+\setminus QQPer$ \, \, \, is the set of \,{\bf hyperhyperprimitive} \\
& & words.  
\end{tabular}
\end{definition}

The different kinds of generalized periodicity are illustrated in the Figures \ref{f3}--\ref{f8}.

\begin{figure}
\vspace*{-12pt}
\begin{center}
  \includegraphics[scale=1.2]{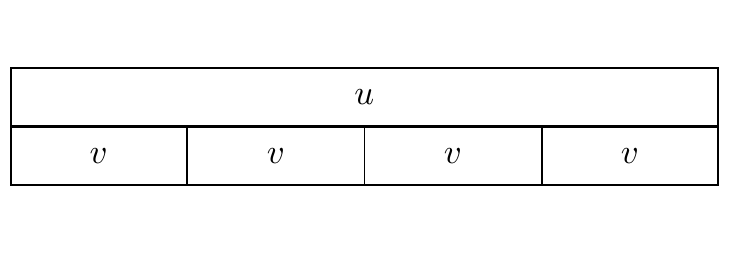}
\end{center}
\vspace*{-24pt}
\caption{$u$ is periodic, $u\in Per$, $v=root(u)$} 
\label{f3}
\end{figure}

\begin{figure}
\vspace*{-24pt}
\begin{center}
  \includegraphics[scale=1.2]{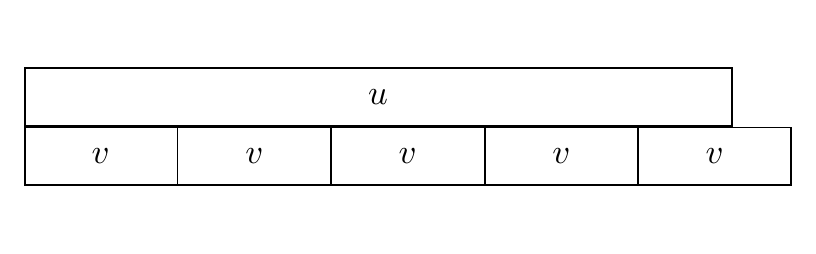}
\end{center}
\vspace*{-24pt}
\caption{$u$ is semi periodic, $u\in SPer$, $v=sroot(u)$} \label{f4}
\end{figure}

\begin{figure}
\vspace*{-24pt}
\begin{center}
   \includegraphics[scale=1.2]{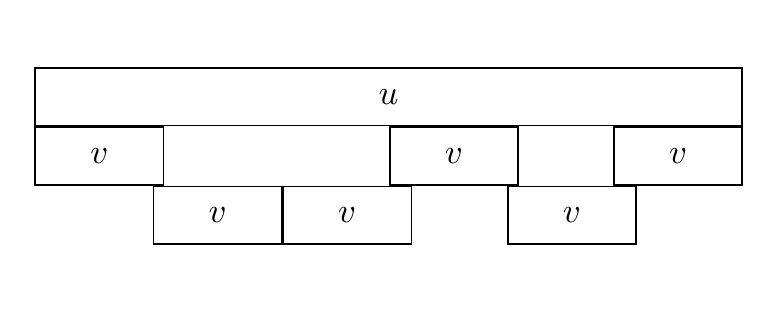}
\end{center}
\vspace*{-36pt}
\caption{$u$ is quasi-periodic, $u\in QPer$, $v=hroot(u)$} \label{f5}
\end{figure}

\begin{figure}
\vspace*{-24pt}
\begin{center}
   \includegraphics[scale=1.2]{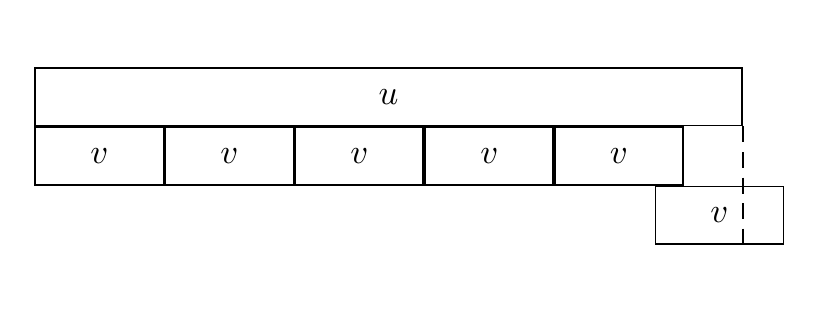}
\end{center}
\vspace*{-36pt}
\caption{$u$ is pre-periodic, $u\in PSPer$, $v=ssroot(u)$} \label{f6}
\end{figure}

\begin{figure}
\vspace*{-36pt}
\begin{center}
   \includegraphics[scale=1.2]{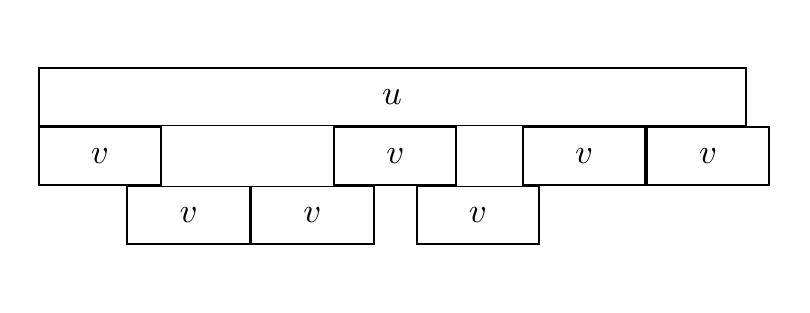}
\end{center}
\vspace*{-36pt}
\caption{$u$ is semi-quasi-periodic, $u\in SQPer$, $v=shroot(u)$} \label{f7}
\end{figure}

\begin{figure}
\vspace*{-36pt}
\begin{center}
   \includegraphics[scale=1.2]{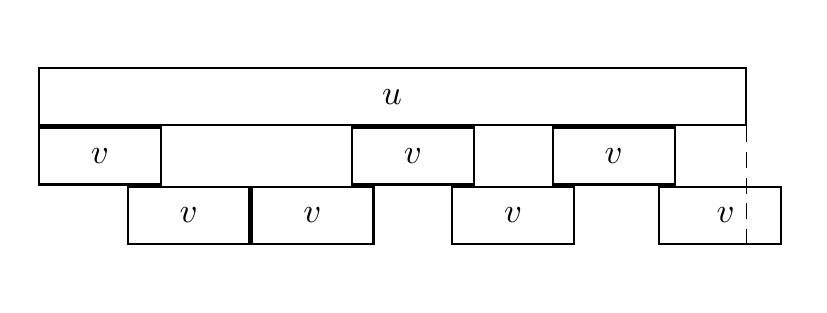}
\end{center}
\vspace*{-35pt}
\caption{$u$ is quasi-quasi-periodic, $u\in QQPer$, $v=hhroot(u)$} \label{f8}
\end{figure}

\begin{figure}
\vspace*{-24pt}
\begin{center}
  \includegraphics[scale=0.9]{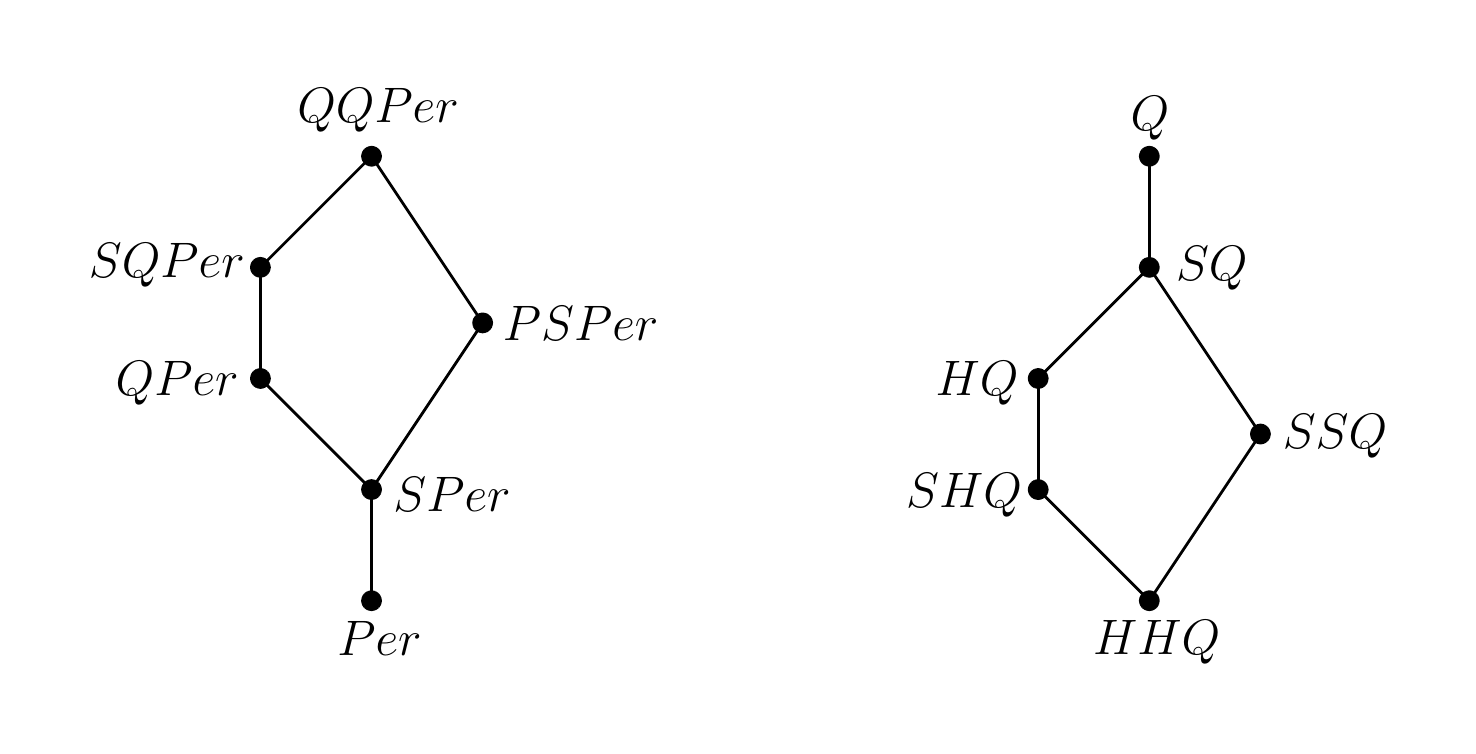}
\end{center}
\vspace*{-24pt}
\caption{Inclusion structure} \label{f9}
\end{figure}

\begin{theorem} \label{t50} The sets from Definition \ref{d49} have the inclusion structure as given in Figure \ref{f9}. The lines in this figure denote strict inclusion from bottom to top. Sets which are not connected by such a line are incomparable under inclusion.
\end{theorem}

\begin{proof} Because of the duality between the sets, it is enough to prove the left structure in Figure \ref{f9}. Let $u\in SPer$, it means, $u=v^nq$ where $n\ge 2$ and $q\sqsubset v$. Thus $v=qr$ for some $r\in \zi ^*$ and $u=(qr)^nq\in (qrq)^{\otimes n}$ and therefore $u\in QPer$ and $SPer\subseteq QPer$. The remaining inclusions are clear by the definition. To show the strictness of the inclusions we can use the following examples: \\
$u_1 = abaababab$, \, $u_2 = aababaababaabaab$, \, $u_3 = aabaaabaaba$, \, $u_4 = abaabab$, \, $u_5 = ababa$. \\
Then \, $u_1\in QQPer\setminus (SQPer\cup PSPer)$, \, $u_2\in SQPer\setminus QPer$, \\
$u_3\in QPer\setminus PSPer$, \, $u_4\in PSPer\setminus SQPer$, \, and \, $u_5\in SPer\setminus Per$. \\
$u_3$ and $u_4$ also prove the incomparability. 
\end{proof} 

The six different kinds of periodicity resp. primitivity of words give rise to define six types of roots where the first one is again that from Definition \ref{d6}.

\begin{definition} \label{d51} Let $u\in \zi ^+$. \\
The shortest word $v$ such that there exists a natural number $n$ with \\ 
$u=v^n$ \, is called the {\bf root} of $u$, denoted by $root(u)$. \\
The shortest word $v$ such that there exists a natural number $n$ with \\
$u\in v^n\cdot Pr(v)$ \, is called the {\bf strong root} of $u$, denoted by $sroot(u)$. \\
The shortest word $v$ such that there exists a natural number $n$ with \\
$u\in v^{\otimes n}$ \, is called the {\bf hyperroot} of $u$, denoted by $hroot(u)$. \\
The shortest word $v$ such that there exists a natural number $n$ with \\
$u\in \{v^n\}\otimes Pr(v)$ \, is called the {\bf super strong root} of $u$, denoted by $ssroot(u)$. \\
The shortest word $v$ such that there exists a natural number $n$ with \\
$u\in v^{\otimes n}\cdot Pr(v)$ \, is called the {\bf strong hyperroot} of $u$, denoted by $shroot(u)$. \\
The shortest word $v$ such that there exists a natural number $n$ with \\
$u\in v^{\otimes n}\otimes Pr(v)$ \, is called the {\bf hyperhyperroot} of $u$, denoted by $hhroot(u)$. \\ 
If $L$ is a language, then \, $root(L) =_{Df} \{root(p):\, p\in L\, \wedge \, p\neq \epsilon \}$ \, is the {\bf root of} $L$. Analogously $sroot(L)$, $hroot(L)$, $ssroot(L)$, $shroot(L)$ and $hhroot(L)$ are defined.
\end{definition}

The six kinds of roots are illustrated in the Figures \ref{f3}--\ref{f8} (if $v$ is the shortest prefix with the appropriate property).

$root$, $sroot$, $hroot$, $ssroot$, $shroot$ and $hhroot$ are word functions over $\zi ^+$, i.e., functions from $\zi ^+$ to $\zi^+$. Generally, for word functions we define the following partial ordering, also denoted by $\sqsubseteq $. \\
$dom(f)$ for a function $f$ denotes the {\bf domain} of $f$.

\begin{definition} \label{d52} For word functions $f$ and $g$ having the same domain, \\
$f\sqsubseteq g \,=_{Df} \,\forall u(u\in dom(f) \rightarrow  f(u)\sqsubseteq g(u))$.
\end{definition}

\begin{figure}
\vspace*{-24pt}
\begin{center}
    \includegraphics[scale=1.2]{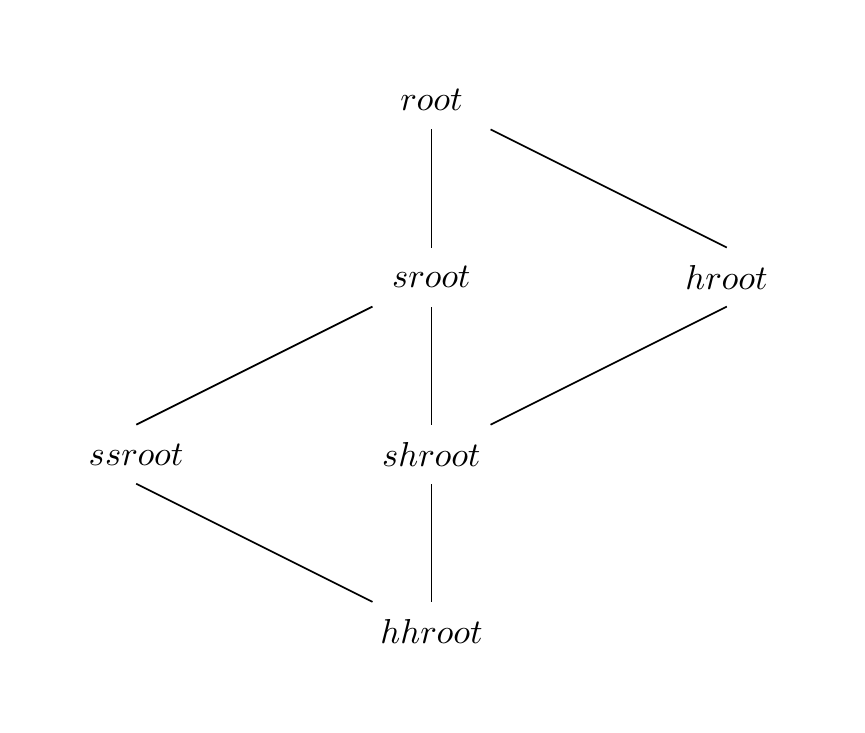}
\end{center}
\vspace*{-24pt}
\caption{Partial ordering of the root-functions} \label{f10}
\end{figure}

\begin{theorem} \label{t53} The partial ordering \, $\sqsubseteq $ \, for the functions from Definition \ref{d51} is given in Figure \ref{f10}.
\end{theorem}

\begin{proof} It follows from the definition, that for an arbitrary word $u\in \zi ^+$ and its roots we have the prefix relationship as shown in the figure. It remains to show the strict prefixes and incomparability. This can be done, for instance, by the following examples. Let \, $u_1 = abaabaababaabaabab$, \, $u_2 = abaabaabab$, \, and \, $u_3 = abaababaabaabaab$. Then \\
$hhroot(u_1) = aba \sqsubset  shroot(u_1) = abaab \sqsubset  ssroot(u_1) = sroot(u_1) = abaabaab \sqsubset  hroot(u_1) = abaabaabab \sqsubset  root(u_1) = u_1$, \\
$ssroot(u_2) = aba \sqsubset  shroot(u_2) = abaab \sqsubset  sroot(u_2) = abaabaab \sqsubset  hroot(u_2) = u_2$, \, and \\
$hroot(u_3) = abaab \sqsubset  sroot(u_3) = abaababaaba$, \, which proves our figure. 
\end{proof}

For most words $u$, some of the six roots coincide, and we have the question how many roots of $u$ are different, and whether there exist words $u$ such that all the six roots of $u$ are different from each other. This last question was raised in \cite{15}, and it was first assumed that they do not exist. But in 2010 Georg Lohmann discovered the first of such words.

\begin{definition} \label{d54} Let $k\in \{1,2,3,4,5,6\}$. A word $u\in \zi ^+$ is called a {\bf $k$-root word} if \\
$|\{root(u),sroot(u),hroot(u),ssroot(u),shroot(u),hhroot(u)\}|=k$. \\
A 6-root word is also called a {\bf Lohmann word}. \\
$u$ is called a {\bf strong $k$-root word} if it is a $k$-root word and $root(u)\neq u$, it means, it is a periodic $k$-root word.
\end{definition}

The following theorems give answers to our questions. The proofs are easy or will be published elsewhere.

\begin{theorem} \label{t55} The lexicographic smallest $k$-root words are \,$a$ \,for $k=1$, \\
$aba$ \,for $k=2$, \,$ababa$ \,for $k=3$, \,$abaabaabab$ \,for $k=4$, \\
$abaabaababaabaabab$ \,for $k=5$, and \\
$ababaabababaababaababababaabab$ \,for $k=6$. \\
The lexicographic smallest strong $k$-root words are \,$aa$ \,for $k=1$, \\
$abaababaab$ \,for $k=2$, \,$(ab^3abab^3abab^3)^2$ \,for $k=3$, and \\
$(ababaabababaabab)^2$ \,for $k=4$.
\end{theorem}

\begin{theorem} \label{t56} There exist no strong $k$-root words for $k=5$ and $k=6$.
\end{theorem}

\begin{theorem} \label{t57} Let $v$ and $w$ be words such that $\epsilon \sqsubset v\sqsubset w$, \,$wv\not\sqsubseteq p^l$ \,for some $p\sqsubset w$ and $l>1$ and $k_1,k_2,k_3$ be natural numbers with $2\le k_1<k_2<k_3\le 2k_1$. Then $u = w^{k_1}vw^{k_2}vw^{k_1}vw^{k_3}vw^{k_3-k_1}$ is a Lohmann word.
\end{theorem}

It is still open whether the sufficient condition in the last theorem is also a necessary condition for Lohmann words.

Let us now examine whether the results from the former sections are also true for generalized periodicity and primitivity. First, we give generalizations of Corollary \ref{c9} and Theorem \ref{t13}. For their proofs we refer to \cite{15}.

\begin{lemma} \label{l58} $w\notin SQ$ if and only if $w=pq=qr$ for some $p,q,r\in \zi ^+$ and $|q|\ge \frac{|w|}{2}$.
\end{lemma}

\begin{lemma} \label{l59} If $aw\notin SQ$ and $wb\notin SQ$, where $w\in \zi ^+$ and $a,b\in \zi $, then $awb\notin SQ$.
\end{lemma}

\begin{lemma} \label{l60} If $aw\notin HQ$ and $wb\notin HQ$, where $w\in \zi ^+$ and $a,b\in \zi $, then $awb\notin HQ$.
\end{lemma}

Theorem \ref{t19} remains true for each of the sets from Definition \ref{d49}. The Theorems \ref{t21}, \ref{t22}, and \ref{t24} with their proofs are passed to each of the languages $SQ$, $HQ$, $SSQ$, $SHQ$, and $HHQ$. Also the non-context-freeness of each of the sets of generalized periodic words is simple as remarked after Theorem \ref{t21}. The context-freeness of the sets of generalized primitive words is open just as that of $Q$.

Using Lemma \ref{l59} and Lemma \ref{l60} it can be shown that Theorem \ref{t27} is also true for $SQ$ and $HQ$. Also none of $SSQ$, $SHQ$, $HHQ$, and the sets of generalized periodic words is a contextual language of any kind.

Theorem \ref{t30} and its proof remain true for each of the sets from Definition \ref{d49}. Theorem \ref{t29} is true for $SQ$ where the proof uses Lemma \ref{l58}. Whether the time bound $n^2$ is also optimal for accepting one of the remaining sets remains open. Theorem \ref{t31} and its proof remain true for each of the roots from Definition \ref{d51}.

\section*{Acknowledgements}

The author is grateful to Antal Iv\'anyi in Budapest for his suggestion to 
write this paper, for Martin H\"unniger in Jena for his help with the 
figures, and to Peter Leupold in Kassel and to the anonymous referee for 
some hints.

This work was supported by the project under the grant agreement no. T\'AMOP 4.2.1/B-09/1/KMR-2010-0003 (E\"otv\"os Lor\'and University, Budapest) financed by the European Union and the European Social Fund.

\rightline{\emph{Received:  December 16,  2010 {\tiny \raisebox{2pt}{$\bullet$\!}} Revised: February 22, 2011}} 

\end{document}